\documentclass[pra,twocolumn,showpacs,superscriptaddress,preprintnumbers,eqsecnum,notitlepage]{revtex4-2}
\usepackage[utf8]{inputenc}
\usepackage[bookmarks=false,colorlinks=true,urlcolor=blue,citecolor=blue,linkcolor=blue]{hyperref}
\usepackage{natbib}
\usepackage{booktabs}
\usepackage{qcircuit}
\usepackage{array}
\usepackage{amsmath,amssymb,mathtools}
\usepackage{graphicx}
\usepackage{bm,color}
\usepackage[capitalise]{cleveref}
\usepackage{multirow}
\usepackage{amsfonts} 
\usepackage{amsmath} 
\usepackage{blkarray}
\usepackage{bbold}
\usepackage{tikz}
\usepackage{graphicx}
\usepackage{changes}
\usepackage[english]{babel}

\begin{document}

\title{Qutrit Circuits and Algebraic Relations:\\ A Pathway to Efficient Spin-1 Hamiltonian Simulation}

\author{Oluwadara Ogunkoya}
\email{ogunkoya@fnal.gov}
\affiliation{Fermi National Accelerator Laboratory, Batavia, IL 60510, USA}

\author{Joonho Kim}
\email{joonho0@gmail.com}
\affiliation{Rigetti Computing, Berkeley, CA 94710, USA}

\author{Bo Peng}
\email{peng398@pnnl.gov}
\affiliation{Physical and Computational Sciences Division, Pacific Northwest National Laboratory, Richland, WA 99352, USA}

\author{A. Barış Özgüler}
\email{ozguler@wisc.edu}
\affiliation{Fermi National Accelerator Laboratory, Batavia, IL 60510, USA}

\author{Yuri Alexeev}
\email{yuri@alcf.anl.gov}
\affiliation{Computational Science Division, Argonne National Laboratory, Lemont, IL 60439, USA}

\date{\today}

\begin{abstract}
Quantum information processing has witnessed significant advancements through the application of qubit-based techniques within universal gate sets. Recently, exploration beyond the qubit paradigm to $d$-dimensional quantum units or qudits has opened new avenues for improving computational efficiency. This paper delves into the qudit-based approach, particularly addressing the challenges presented in the high-fidelity implementation of qudit-based circuits due to increased complexity. As an innovative approach towards enhancing qudit circuit fidelity, we explore algebraic relations, such as the Yang-Baxter-like turnover equation, which may enable circuit compression and optimization. The paper introduces the turnover relation for the three-qutrit time propagator and its potential use in reducing circuit depth. We further investigate whether this relation can be generalized for higher-dimensional quantum circuits, including a focused study on the one-dimensional spin-1 Heisenberg model. Our work outlines both rigorous and numerically efficient approaches to potentially achieve this generalization, providing a foundation for further explorations in the field of qudit-based quantum computing.
\end{abstract}

\maketitle

\section{Introduction}

Quantum information processing through a gate-based quantum computing approach with qubits involves a universal gate set consisting of single-qubit gates in the $SU(2)$ group and entangling two-qubit gates~\cite{DiVincenzo1995TwoBit}. This approach has been intensively studied in recent years for applications in quantum information science. Particularly in the quantum error correction (QEC), the qubit-based surface code~\cite{bravyi1998quantum,Dennis2002Topological} has been thus far the primary route for error detection and correction~\cite{Chen2021Exponential,Marques2021Logical,Krinner2022Realizing,Zhao2022Realization}.

\textcolor{black}{Nevertheless, for certain specific applications, it has been discussed that a more generalized $d$-dimensional ($d>2$) quantum unit, or qudit, might offer advantages over the qubit system. This is because the qudit-based approach allows exploration beyond two levels, potentially enhancing performance through access to a larger computational space and requiring fewer entangling gates for certain algorithms. qudits have theoretically exhibited advantages through compact logical encoding (to overcome erasure and ternary errors)~\cite{Kapit2016Hardware,Muralidharan2017Overcoming,Majumdar2018Quantum}), as well as enhancements in efficiency and fault tolerance~\cite{Campbell2012Magic,Campbell2014Enhanced,otten2020impacts, alam2022quantum, ozguler2022dynamics}.
In the field of quantum key distribution (QKD), studies suggest that qudits can increase the average raw key rates and improve robustness and reliability~\cite{PhysRevLett.96.090501,PhysRevA.67.012311}.
Regarding other applications, theoretical reports have proposed the use of qutrits, the simplest qudit system, to enhance quantum algorithms (e.g., Shor's factoring~\cite{Bocharov2017Factoring}, Grover's search~\cite{Gokhale2019Asymptotic,Bullock2005Asymptotically}, quantum Fourier transformation~\cite{Pavlidis2021Quantum}), quantum simulations~\cite{gustafson2022noise}, quantum cryptography~\cite{Bechmann-Pasquinucci2000Quantum,Brus2002Optimal}, and quantum communication~\cite{Vaziri2002Experimental}.}

\textcolor{black}{Discussions have also been extended to more fundamental problems, such as the Byzantine Agreement~\cite{PhysRevLett.87.217901}, efficient Toffoli gates~\cite{PhysRevA.75.022313}, and quantum channels demonstrating the superadditivity of classical capacity~\cite{PhysRevLett.90.167906}.} 
\textcolor{black}{Certain quantum gates, such as parameterized gates for quantum heuristics~\cite{li5benchmarking}, can be implemented more naturally in qutrit systems~\cite{ozguler2022numerical}, as qutrits offer a more direct and efficient mapping for spin-1 models compared to traditional qubits. This direct mapping is particularly beneficial in studying phenomena like the Haldane gap and many-body localization in spin-1 chains~\cite{ozguler2018steering, ozguler2019response, ozguler2021excitation, cao2021speedup}. The approach in Ref.~\cite{cao2021speedup}  involves catalyzing the algorithm so that its evolution mimics a Heisenberg model in a delocalized phase, which demonstrates a speedup in finding the ground state of the random-field Ising model due to gap amplification, with promising scalability indications.}

\textcolor{black}{It's worth noting that the advantages mentioned above can be offset by the costs of implementing and operating qudits in real quantum applications. } 
The universal gate set for qudits consists of single-qudit gates in the $SU(d)$ group and entangling two-qudit gates~\cite{Muthukrishnan2000Multivalued,Zhou2003Quantum,Brennen2005Criteria}. Recent efforts have demonstrated that the universal qudit gate set and their coherent control can be implemented in superconducting transmons~\cite{Blok2021Quantum,Goss2022High,fischer2022universal}, photonic circuits~\cite{Chi2022programmable}, and trapped ions~\cite{Ringbauer2022Universal,hrmo2022native}. However, implementing high-fidelity qudit-based circuits also presents challenges, mainly due to the increased complexity in the design, fabrication, and control of quantum systems with higher dimensions.

Towards improving the fidelity of qudit circuit, one direction could be to explore some algebraic relations between qudit circuits for the purpose of circuit compression and optimization, which would result in more fault-resilient performance. Similar exploration for the qubit-based circuits have recently become quite active~\cite{bassman2021constantdepth,Kokcu2022Algebraic,Camps2021Algebraic,lin2021real,cirstoiu2020variational,atia2017fast,berthusen2022quantum,barratt2021parallel}. For example, the Yang-Baxter equation (YBE), which was originally introduced in theoretical physics~\cite{yang1967some} and statistical mechanics~\cite{baxter2016exactly}, has recently been shown to have connections to topological entanglement, quantum entanglement, and quantum computational universality~\cite{ge2016YBE,nayak2008nonAbelian,kauffman2010topological,zhang2013integrable,vind2016experimental,batchelor2016YB,Peng2022Quantum,gulania2022quybe}. 
\textcolor{black}{For example, proposals have been reported for efficiently checking YBE in quantum devices~\cite{Wang2020Experimental,Zheng2013Direct,vind2016experimental}.}
In our previous work~\cite{Peng2022Quantum}, we proved that for some model systems, the two-qubit time propagator $\mathcal{R}_{\theta,\delta}$, parametrized by a rotation angle $\theta$ and a phase $\delta$, bears a similar algebraic form to the $SU(2)$ solution of the YBE. Therefore, the turn-over relationship (\ref{YBE_general}) can hold as long as certain algebraic relations between the parameters on both sides are satisfied. Remarkably, this turn-over relation can be utilized to compress the corresponding time evolution circuit to a depth that scales linearly with respect to the number of qubits. 

\begin{equation}
    \resizebox{0.85\hsize}{!}{%
    $\Qcircuit @C=.5em @R=.5em {
    & \multigate{1}{\mathcal{R}_1(\boldsymbol\alpha)} & \qw & \multigate{1}{\mathcal{R}_3(\boldsymbol\gamma)} & \qw &\\
    & \ghost{\mathcal{R}_1(\boldsymbol\alpha)} & \multigate{1}{\mathcal{R}_2(\boldsymbol\beta)} & \ghost{\mathcal{R}_3(\boldsymbol\gamma)} & \qw & ~~~ = \\
    & \qw & \ghost{\mathcal{R}_2(\boldsymbol\beta)} & \qw & \qw &}
    ~~~~
    \Qcircuit @C=.5em @R=.5em {
    & \qw & \multigate{1}{\mathcal{R}_5(\boldsymbol\epsilon)} & \qw & \qw \\
    & \multigate{1}{\mathcal{R}_4(\boldsymbol\delta)} & \ghost{\mathcal{R}_5(\boldsymbol\epsilon)} & \multigate{1}{\mathcal{R}_6(\boldsymbol\zeta)} & \qw \\
    & \ghost{\mathcal{R}_4(\boldsymbol\delta)} & \qw & \ghost{\mathcal{R}_6(\boldsymbol\zeta)} & \qw }$
} \label{YBE_general}
\end{equation}

This observation then opens the question of whether these turn-over relations can be generalized for quantum circuits with higher dimensions. Mathematically, there have been discussions on finding the high-dimension solutions to the generalized YBE~\cite{Rowell2010Extraspecial,galindo2011generalized,rowell2010quaternionic,Chen2012Generalized,isaev2022lectures}; however, numerical searching of these high-dimension solutions can be challenging. So far, in addition to the $SU(2)$ solutions, only an $8\times 8$ solution to a generalized YBE has been reported and used to generate the Greenberger-Horne-Zeillinger states~\cite{Rowell2010Extraspecial}. In this paper, as an exploratory effort in this direction, we primarily focus on establishing a similar turn-over relation that can be utilized for the efficient quantum simulation of the quantum time dynamics of the one-dimensional spin-1 Heisenberg model. In particular, 
\textcolor{black}{a natural mapping of the spin-1 system's states onto the qutrit states leading to more straightforward or efficient quantum simulations allow us to efficiently}
search for (1) the existence of rigorous algebraic conditions for the similar turn-over relations to hold, and (2) a numerically efficient approach that can provide  imprecise but sufficiently accurate turn-over circuits in the absence of rigorous algebraic relations.

In the following sections, we  first define some notations that will be used in this paper. Then, we show for some simple models, rigorous turn-over relations do exist. Finally, for models without rigorous turn-over relations, we propose a numerical recipe to achieve inexact but accurate enough qutrit circuit substitutes. The numerical recipe and the corresponding error analysis are provided for the three-qutrit circuit simulating the time dynamics of a three-site spin-1 Heisenberg model. We conclude this work by offering some remarks on our future effort.

\section{Notations and Spin Algebra}

The closed-system dynamics of a one-dimensional array of level-$d$ variables is realized by $U(d^N)$ unitary matrices, where $N$ denotes the size of the system. Throughout this paper, however, we specifically treat qudits as spin $s=(d-1)/2$ quantum states and consider their time evolution with certain bilinear spin Hamiltonians. 

We recall that for $d=2$, $s=1/2$ the spin operators, satisfying the $SU(2)$ commutation algebra
\begin{align}
    [S^x, S^y] = iS^z,~~ [S^y, S^z] = iS^x,~~ [S^z,S^x] = iS^y,
\end{align}
are halves of the Pauli matrices:
\begin{align}
    X = \left( \begin{array}{cc} 0 &  1 \\ 1 &  0 \end{array} \right),~~
    Y = \left( \begin{array}{cc} 0 & -i \\ i &  0 \end{array} \right),~~
    Z = \left( \begin{array}{cc} 1 &  0 \\ 0 & -1 \end{array} \right).
\end{align}
The $SU(2)$ algebra allows a  quadratic Casimir invariant,
\begin{align}
    (S^x)^2 + (S^y)^2 + (S^z)^2 =  s(s+1) \mathbf{1}. 
\end{align}

%
For three-level systems ($d=3$, $s=1$) the $z$-basis representation of the spin-$1$ operators becomes   
\begin{align}
    S^x &= \frac{1}{\sqrt{2}}\begin{pmatrix}0 & 1 & 0 \\ 1 & 0 & 1 \\ 0 & 1 & 0\end{pmatrix} 
         = \frac{1}{\sqrt{2}}\left( X \oplus 0 + 0 \oplus X\right), \label{eq:qutrit_x} \\ 
    S^y &= \frac{1}{\sqrt{2}}\begin{pmatrix}0 & -i & 0 \\ i & 0 & -i \\ 0 & i & 0\end{pmatrix}  
         = \frac{1}{\sqrt{2}}\left( Y \oplus 0 + 0 \oplus Y\right), \label{eq:qutrit_y} \\
    S^z &= \begin{pmatrix}1 & 0 & 0 \\ 0 & 0 & 0 \\ 0 & 0 & -1\end{pmatrix} 
         = \left( Z \oplus 0 + 0 \oplus Z\right). \label{eq:qutrit_z} 
\end{align}
As an alternative to (\ref{eq:qutrit_x}), (\ref{eq:qutrit_y}), and (\ref{eq:qutrit_z}), it is sometimes more convenient to use the adjoint representation of the spin operators,
\footnote{\textcolor{black}{It's something interesting to see how to implement these spin-1 operators using spin-1/2 matrices. Take Eq. (2.7) as an example, we can expand it to a $4\times 4$ matrix and represent using spin-1/2 Pauli matrices, 
\begin{align}
    \tilde{S}^x &= \left(\begin{array}{ccc} 0 & 0 & 0 \\  0 & 0 & i \\  0 & -i & 0\end{array}\right) \Rightarrow \left(\begin{array}{cccc} 0 & 0 & 0 & 0 \\  0 & 0 & i & 0 \\  0 & -i & 0 & 0 \\ 0 & 0 & 0 & 0 \end{array}\right) = I_2 \otimes A - A \otimes I_2
\end{align}
where $A = \frac{1}{2}(-i X_2 + Y_2)$, and $I_2, X_2, Y_2$ are spin-1/2 Pauli matrices
\begin{align}
    I_2 = \left(\begin{array}{cc} 1 & 0 \\  0 & 1 \end{array}\right),~~ 
    X_2 = \left(\begin{array}{cc} 0 & 1 \\  1 & 0 \end{array}\right),~~ 
    Y_2 = \left(\begin{array}{cc} 0 & -i \\  i & 0 \end{array}\right).
\end{align}}}
\begin{align}
     \tilde{S}^x &= \begin{pmatrix}0 & 0 & 0 \\  0 & 0 & i \\  0 & -i & 0\end{pmatrix} = 0 \oplus (-Y), \label{sx_adj} \\
     \tilde{S}^y &= \begin{pmatrix}0 & 0 & i \\  0 & 0 & 0 \\ -i &  0 & 0\end{pmatrix} 
           = P_y \tilde{S}^x P_y^\dagger, \label{sy_adj} \\
     \tilde{S}^z &= \begin{pmatrix}0 & i & 0 \\ -i & 0 & 0 \\  0 &  0 & 0\end{pmatrix} 
           = P_z \tilde{S}^x P_z^\dagger \label{sz_adj}
\end{align}
with permutation matrices $P_y$ and $P_z$ given by
\begin{align}
    P_y = \begin{pmatrix}0&1&0\\ 1&0&0\\ 0&0&1 \end{pmatrix},~~
    P_z = \begin{pmatrix}0&1&0\\ 0&0&1\\ 1&0&0 \end{pmatrix},
\end{align}
and
\begin{align}
    (\tilde{S}^x)^2 &= \text{diag}(0,1,1),\\
    (\tilde{S}^y)^2 &= 
    \text{diag}(1,0,1),\\
    (\tilde{S}^z)^2 &= 
    \text{diag}(1,1,0).
\end{align}
Similar to the $SU(2)$ cases, $\{\tilde{S}^x,\tilde{S}^y,\tilde{S}^z\}$ follows the commutation algebra
\begin{align}
    [\tilde{S}^x, \tilde{S}^y] = i\tilde{S}^z,~~ [\tilde{S}^y, \tilde{S}^z] = i\tilde{S}^x,~~ [\tilde{S}^z, \tilde{S}^x] = i\tilde{S}^y
\end{align}
The two representations $\{S^x,S^y,S^z\}$ and $\{\tilde{S}^x,\tilde{S}^y,\tilde{S}^z\}$ are connected through the basis change between spherical and Cartesian coordinates, 
\begin{align}
\label{eq:qutrit-basis-change}
    \tilde{S}^a = U_{+} S^a U^\dagger_{+},~~ a = x, y, z  
\end{align}
with
\begin{align}
U_\pm = \frac{1}{\sqrt{2}}\begin{pmatrix}
-1 & 0 & \pm 1 \\ 
+i & 0 & \pm i \\ 
0 & \sqrt{2} & 0
\end{pmatrix}, \label{mat_U}
\end{align}
The basis change \eqref{eq:qutrit-basis-change} does not affect the algebraic relations and the circuit substitutes established in the following sections.

Some algebraic relations of the spin-1 operators are worth mentioning. For example, for $n\ge 1$, we have
\begin{align}
\begin{cases}
    (\tilde{S}^a)^{2n} = (\tilde{S}^a)^2\\ (\tilde{S}^a)^{2n+1} = \tilde{S}^a \\ \tilde{S}^a\tilde{S}^b\tilde{S}^a = \mathbf{0}_3
\end{cases} \quad \text{for } a\neq b\in \{x,y,z\}.
\end{align}
which implies
\begin{align}
\mathcal{U}_x(\alpha) &= \exp{(-i \alpha \tilde{S}^x \otimes \tilde{S}^x)} \label{eq:Ux} \\
    &= \mathbf{I}_9 - i\sin(\alpha)(\tilde{S}^x \otimes \tilde{S}^x) - 2\sin^2(\frac{\alpha}{2})(\tilde{S}^x \otimes \tilde{S}^x)^2  \notag \\
    &= \begin{pmatrix} 
        \mathbf{I}_3 & \mathbf{0}_3 & \mathbf{0}_3 \\
        \mathbf{0}_3 & \mathbf{I}_3- 2(\sin(\frac{\alpha}{2})\tilde{S}^x)^2 & \sin(\alpha)\tilde{S}^x \\
        \mathbf{0}_3 & -\sin(\alpha)\tilde{S}^x & \mathbf{I}_3- 2(\sin(\frac{\alpha}{2})\tilde{S}^x)^2 
    \end{pmatrix}\notag,
\end{align}
\begin{align}    
    \mathcal{U}_y(\alpha) &= \exp{(-i \alpha \tilde{S}^y \otimes \tilde{S}^y)} \label{eq:Uy} \\
    &= \mathbf{I}_9 - i\sin(\alpha)(\tilde{S}^y \otimes \tilde{S}^y) - 2\sin^2(\frac{\alpha}{2})(\tilde{S}^y \otimes \tilde{S}^y)^2  \notag \\
    &= \begin{pmatrix} 
        \mathbf{I}_3- 2(\sin(\frac{\alpha}{2})\tilde{S}^y)^2 & \mathbf{0}_3 & \sin(\alpha)\tilde{S}^y \\
        \mathbf{0}_3 & \mathbf{I}_3 & \mathbf{0}_3 \\
        -\sin(\alpha)\tilde{S}^y & \mathbf{0}_3 & \mathbf{I}_3- 2(\sin(\frac{\alpha}{2})\tilde{S}^y)^2 
    \end{pmatrix}\notag,
\end{align}
\begin{align}
    \mathcal{U}_z(\alpha) &= \exp{(-i \alpha \tilde{S}^z \otimes \tilde{S}^z)} \label{eq:Uz} \\
    &= \mathbf{I}_9 - i\sin(\alpha)(\tilde{S}^z \otimes \tilde{S}^z) - 2\sin^2(\frac{\alpha}{2})(\tilde{S}^z \otimes \tilde{S}^z)^2  \notag \\
    &= \begin{pmatrix} 
        \mathbf{I}_3 - 2(\sin(\frac{\alpha}{2})\tilde{S}^z)^2 & \sin(\alpha)\tilde{S}^z & \mathbf{0}_3 \\
        -\sin(\alpha)\tilde{S}^z & \mathbf{I}_3 - 2(\sin(\frac{\alpha}{2})\tilde{S}^z)^2 & \mathbf{0}_3 \\
        \mathbf{0}_3 & \mathbf{0}_3 & \mathbf{I}_3
    \end{pmatrix}, \notag
\end{align}
where $\mathbf{I}_m$ denotes an $m\times m$ identity matrix and $\mathbf{0}_m$ denotes an $m\times m$ zero matrix. Notably, from (\ref{sy_adj}) and (\ref{sz_adj})
\begin{align}
    &\Qcircuit @C=.5em @R=.5em {
    & \multigate{2}{\mathcal{U}_y(\alpha)} & \qw &\\
    & & & ~~ = \\
    & \ghost{\mathcal{U}_y(\alpha)} & \qw &}
    ~~~~\Qcircuit @C=.5em @R=.5em {
    & \gate{P_y} & \multigate{1}{\mathcal{U}_x(\alpha)} & \gate{P_y{{{}^\dagger}}} & \qw \\
    & \gate{P_y} & \ghost{\mathcal{U}_x(\alpha)} & \gate{P_y{{{}^\dagger}}} & \qw } \label{Uy} \\
    &\Qcircuit @C=.5em @R=.5em {
    & \multigate{2}{\mathcal{U}_z(\alpha)} & \qw & \\
    & & & ~~ = \\
    & \ghost{\mathcal{U}_z(\alpha)} & \qw & }
    ~~~~\Qcircuit @C=.5em @R=.5em {
    & \gate{P_z} & \multigate{1}{\mathcal{U}_x(\alpha)} & \gate{P_z{{{}^\dagger}}} & \qw \\
    & \gate{P_z} & \ghost{\mathcal{U}_x(\alpha)} & \gate{P_z{{{}^\dagger}}} & \qw } \label{Uz}
\end{align}
A more interesting feature of $\mathcal{U}_{a}(\alpha)$ with $a\in\{x,y,z\}$ is that
\begin{align}
    &\Qcircuit @C=.5em @R=.5em {
    & \multigate{1}{\mathcal{U}_a(\alpha)} & \qw & \qw &\\
    & \ghost{\mathcal{U}_a(\alpha)} & \multigate{1}{\mathcal{U}_a(\alpha)} & \qw & ~~= \\
    & \qw & \ghost{\mathcal{U}_a(\alpha)} & \qw &}
    ~~~~\Qcircuit @C=.5em @R=.5em {
    & \qw & \multigate{1}{\mathcal{U}_a(\alpha)} & \qw \\
    & \multigate{1}{\mathcal{U}_a(\alpha)} & \ghost{\mathcal{U}_a(\alpha)} & \qw \\
    & \ghost{\mathcal{U}_a(\alpha)} & \qw & \qw } \label{turnover1}
\end{align}

\section{Yang-Baxter-like Relations in Qutrit Circuit}

This section aims to search Yang-Baxter-type identities for qutrit circuits. Specifically, this means establishing a $(3^3 \times 3^3)$ matrix relation of the following type:
\begin{align}
\begin{split}
    &\big(\mathcal{R}_1(\alpha) \otimes \mathbf{I}_3\big)\big(\mathbf{I}_3 \otimes \mathcal{R}_2(\beta)\big)\big(\mathcal{R}_3(\gamma) \otimes \mathbf{I}_3\big) \\ &~~~~~~~~=\big(\mathbf{I}_3 \otimes \mathcal{R}_4(\delta)\big)\big(\mathcal{R}_5(\epsilon) \otimes \mathbf{I}_3\big)\big(\mathbf{I}_3 \otimes \mathcal{R}_6(\zeta)\big)
\end{split}
\label{eq:ybe-qutrit}
\end{align}
where the Greek letters $\alpha, \cdots, \zeta$ collectively denote continuous rotations that parameterize two-qutrit operators $\mathcal{R}_n$, defined as a product of $\mathcal{U}_a$'s ($a\in \{x, y, z\}$). The rotation angles on the LHS of \eqref{eq:ybe-qutrit} are unrestricted; we require that for all values of $\alpha$, $\beta$, $\gamma$, there should be a value of $\delta$, $\epsilon$, $\zeta$ that satisfies \eqref{eq:ybe-qutrit}. The parameters are typically related via unclosed expressions involving trigonometric functions, derived from element-wise equalities of \eqref{eq:ybe-qutrit}. In the following subsections, we describe analytical and numerical methods to establish \eqref{eq:ybe-qutrit} with various $\mathcal{R}_n$'s.

\subsection{Simple turn-over identities}\label{simple_relation}

We start from the simple case where $\mathcal{R}_n = \mathcal{U}_a$ for $a\in \{x, y, z\}$. If all the rotations are same, then by directly applying (\ref{turnover1}) to the LHS or RHS of (\ref{eq:ybe-qutrit}), we obtained the following YBEs relationships:\\

\noindent $\bullet$ LHS:
\begin{align}
    \resizebox{\hsize}{!}{%
    $\Qcircuit @C=.5em @R=.5em {
    & \multigate{1}{\mathcal{U}_a(\alpha)} & \qw & \multigate{1}{\mathcal{U}_a(\alpha)} & \qw &\\
    & \ghost{\mathcal{U}_a(\alpha)} & \multigate{1}{\mathcal{U}_a(\alpha)} & \ghost{\mathcal{U}_a(\alpha)} & \qw & ~~= \\
    & \qw & \ghost{\mathcal{U}_a(\alpha)} & \qw & \qw &}
    ~~~~\Qcircuit @C=.5em @R=.5em {
    & \multigate{1}{\mathcal{U}_a(2\alpha)} & \qw & \qw &\\
    & \ghost{\mathcal{U}_a(2\alpha)} & \multigate{1}{\mathcal{U}_a(\alpha)} & \qw & ~~= \\
    & \qw & \ghost{\mathcal{U}_a(\alpha)} & \qw &}
    ~~~~\Qcircuit @C=.5em @R=.5em {
    & \qw & \multigate{1}{\mathcal{U}_a(2\alpha)} & \qw \\
    & \multigate{1}{\mathcal{U}_a(\alpha)} & \ghost{\mathcal{U}_a(2\alpha)} & \qw \\
    & \ghost{\mathcal{U}_a(\alpha)} & \qw & \qw }$
} \notag
\end{align}

\noindent $\bullet$ RHS:
\begin{align}
    \resizebox{\hsize}{!}{%
    $\Qcircuit @C=.5em @R=.5em {
    & \qw & \multigate{1}{\mathcal{U}_a(\alpha)} & \qw & \qw &\\
    & \multigate{1}{\mathcal{U}_a(\alpha)} & \ghost{\mathcal{U}_a(\alpha)} & \multigate{1}{\mathcal{U}_a(\alpha)} & \qw & ~~= \\
    & \ghost{\mathcal{U}_a(\alpha)} & \qw & \ghost{\mathcal{U}_a(\alpha)} & \qw &}
    ~~~~\Qcircuit @C=.5em @R=.5em {
    & \multigate{1}{\mathcal{U}_a(\alpha)} & \qw & \qw &\\
    & \ghost{\mathcal{U}_a(\alpha)} & \multigate{1}{\mathcal{U}_a(2\alpha)} & \qw & ~~= \\
    & \qw & \ghost{\mathcal{U}_a(2\alpha)} & \qw &}
    ~~~~\Qcircuit @C=.5em @R=.5em {
    & \qw & \multigate{1}{\mathcal{U}_a(\alpha)} & \qw \\
    & \multigate{1}{\mathcal{U}_a(2\alpha)} & \ghost{\mathcal{U}_a(\alpha)} & \qw \\
    & \ghost{\mathcal{U}_a(2\alpha)} & \qw & \qw }$
} \notag
\end{align}

\noindent $\bullet$ LHS $=$ RHS:
\begin{align}
    \resizebox{\hsize}{!}{%
    $\Qcircuit @C=.5em @R=.5em {
    & \multigate{1}{\mathcal{U}_a(\alpha)} & \qw & \multigate{1}{\mathcal{U}_a(\alpha)} & \qw &\\
    & \ghost{\mathcal{U}_a(\alpha)} & \multigate{1}{\mathcal{U}_a(2\alpha)} & \ghost{\mathcal{U}_a(\alpha)} & \qw & ~~= \\
    & \qw & \ghost{\mathcal{U}_a(2\alpha)} & \qw & \qw &}
    ~~~~\Qcircuit @C=.5em @R=.5em {
    & \qw & \multigate{1}{\mathcal{U}_a(2\alpha)} & \qw & \qw \\
    & \multigate{1}{\mathcal{U}_a(\alpha)} & \ghost{\mathcal{U}_a(2\alpha)} & \multigate{1}{\mathcal{U}_a(\alpha)} & \qw  \\
    & \ghost{\mathcal{U}_a(\alpha)} & \qw & \ghost{\mathcal{U}_a(\alpha)} & \qw }$
} \notag
\end{align}

For more general cases where the rotations are not necessarily same, as explicitly shown in Appendix \ref{app_A}, as long as the following conditions are satisfied
\begin{align}
    \label{eq:param}
    \bigg\{\begin{array}{l}
    \alpha + \gamma = \epsilon + 2k\pi  \\
    \delta + \zeta = \beta + 2k\pi 
    \end{array},
    ~~ k\in \mathbf{Z},
\end{align}
the identity \eqref{eq:ybe-qutrit} holds for all $\mathcal{U}_a$ $(a = x, y, z)$. 
This can also be understood as an immediate consequence of the qubit relations. See Appendix~\ref{sec:qubit_qutrit_identity} for the explanation.

The bonus identities can also be obtained from a simple observation: The spin matrices expressed in the adjoint representation also satisfy, for example,
\begin{align}\label{eq:Sy}
U^\dagger_\pm \tilde{S}^y U_\pm &= \tfrac{1}{\sqrt{2}}(-\tilde{S}^z \mp \tilde{S}^x),
\end{align}
Other cyclic relations can further be obtained by utilizing (\ref{sy_adj}) and (\ref{sz_adj}).
It can be shown that
\begin{align}
V^\dagger_\pm \tilde{S}^x V_\pm &= \tfrac{1}{\sqrt{2}}(-\tilde{S}^y \pm \tilde{S}^z),\label{eq:Sx}\\
W^\dagger_\pm \tilde{S}^z W_\pm &= \tfrac{1}{\sqrt{2}}(+\tilde{S}^x \mp \tilde{S}^y).
\end{align}
with
\begin{align}
V_\pm &= \frac{1}{\sqrt{2}}\begin{pmatrix}
 0 & +i & \pm i \\
 0 & \mp 1 & + 1 \\
 \sqrt{2} & 0 & 0 \\
\end{pmatrix}, \label{mat_V}\\ 
W_\pm &= \frac{1}{\sqrt{2}}\begin{pmatrix}
 \mp 1 & + 1 & 0 \\
 0 & 0 & \sqrt{2} \\
 +i & \pm i & 0 \\
\end{pmatrix}. \label{mat_W}
\end{align}
 Since the Yang-Baxter identity \eqref{eq:ybe-qutrit} is independent of the basis, it implies that $\mathcal{R}_n(\theta) \equiv \exp(i\theta (S^a \pm S^b) \otimes (S^a \pm S^b))$ with any $a \neq b \in \{x,y,z\}$ also satisfies \eqref{eq:ybe-qutrit}.

In the same way, we can derive the identity \eqref{eq:ybe-qutrit} for the following set of extended operators: $\mathcal{R}_n(\theta) = \exp(i\theta (S^x \pm S^y \pm S^z) \otimes (S^x \pm S^y \pm S^z)))$ with any choice of $\pm$ factors. This starts with observing the conjugation relation,
\begin{align}
M^\dagger_{\pm\pm}\, \tilde{S}^z \, M_{\pm\pm} &= \tfrac{1}{\sqrt 3}(\tilde{S}^x \mp \tilde{S}^y \pm \tilde{S}^z),
\end{align}
where 
\begin{align}
\label{eq:conj-matrix-3}
M_{s_1s_2} = \frac{1}{\sqrt{6}}\begin{pmatrix}
 -s_1\, \sqrt{3}i & \sqrt{3} i & 0 \\
 -s_2\, i  & -s_1 s_2 \,i & 2 i \\
 s_2\sqrt{2} & s_1s_2\sqrt{2} & \sqrt{2} \\
\end{pmatrix}
\end{align}
for $s_1, s_2 \in \{\pm 1\}$. Note that the unitarity holds for $M_{\pm\pm}$, 
as for the other matrices in (\ref{mat_U}), (\ref{mat_V}), and (\ref{mat_W}). 
Thus, any fixed combination of $(S^x \pm S^y \pm S^z)$  can replace all $S^z$ that appear in the circuit identity \eqref{eq:ybe-qutrit} through $\mathcal{R}_n = \mathcal{U}_z$ if we insert the resolution of identity
\begin{align*}
    M_{\pm\pm}M_{\pm\pm}^\dagger = \mathbf{I}_3
\end{align*}
wherever needed.

The discussion so far has established the Yang-Baxter-type circuit relations \eqref{eq:ybe-qutrit} for the simple cases
where it is assumed that all $\mathcal{R}$ operators have the same form 
\begin{align}
    \mathcal{R}_n = \mathcal{U}_{s_xs_ys_z}
\end{align}
and depend on a single continuous parameter, i.e.,
\begin{gather}
    \mathcal{U}_{s_xs_ys_z} (\alpha) \equiv \exp{\Big(-i\alpha(\textstyle\sum_{a \in \{x,y,z\}} s_a S^a)^{\otimes 2} \Big)} 
\end{gather}
where $s_x, s_y, s_z \in \{+1, 0, -1\}$. 


\subsection{Numerical methods for approximate identities}
\label{ssec:approx_identities}
\begin{figure*}

{
\setlength{\abovedisplayskip}{0pt}
\setlength{\belowdisplayskip}{0pt}
\setlength{\abovedisplayshortskip}{0pt}
\setlength{\belowdisplayshortskip}{0pt}
\begin{tabular}{@{}p{0.7\linewidth}@{}@{}p{0.2\linewidth}@{}p{0pt}}  
\toprule
\vspace{-0.5\baselineskip}
\begin{center}
Unitary Pairs on Test ($W_L \rightarrow W_R$)
\end{center}
&
\vspace{-0.5\baselineskip}
\begin{center}
\centering Description
\end{center}
\\\toprule
\begin{align*}
\Qcircuit @C=.5em @R=.5em{
    & \multigate{1}{\mathcal{U}_x(\theta)} & \qw & \multigate{1}{\mathcal{U}_y(\theta)}& \qw & \qw & \\
    & \ghost{\mathcal{U}_a(\alpha)} & \multigate{1}{\mathcal{U}_x(\theta)} & \ghost{\mathcal{U}_a(\alpha)} & \multigate{1}{\mathcal{U}_y(\theta)} & \qw & \hspace{0.5cm} \longrightarrow \\
    & \qw & \ghost{\mathcal{U}_a(\alpha)} & \qw & \ghost{\mathcal{U}_a(\alpha)}& \qw & \\ } 
~~\hspace{0.5cm}\Qcircuit @C=.5em @R=.5em{
     & \qw & \multigate{1}{\mathcal{U}_y(\zeta)}& \qw & \multigate{1}{\mathcal{U}_x(\mu)} & \qw\\
    & \multigate{1}{\mathcal{U}_y(\lambda)} & \ghost{\mathcal{U}_a(\alpha)} & \multigate{1}{\mathcal{U}_x(\sigma)} & \ghost{\mathcal{U}_a(\alpha)} & \qw\\
    &  \ghost{\mathcal{U}_a(\alpha)} & \qw & \ghost{\mathcal{U}_a(\alpha)}& \qw & \qw\\ } \\
\end{align*}
\vspace{-\baselineskip}
& 
\vspace{\baselineskip}
\centering Trotter scheme $T_1$ \\[0.1em] \eqref{eq:trot1} 
& \\
\midrule
\begin{align*}
\Qcircuit @C=.5em @R=.5em{
    & \multigate{1}{\mathcal{U}_x(\theta)} & \multigate{1}{\mathcal{U}_y(\theta)} & \qw & \qw & \qw & \\
    & \ghost{\mathcal{U}_a(\alpha)} & \ghost{\mathcal{U}_a(\alpha)} & \multigate{1}{\mathcal{U}_x(\theta)} & \multigate{1}{\mathcal{U}_y(\theta)}  & \qw & \hspace{0.5cm} \longrightarrow \\
    & \qw & \qw & \ghost{\mathcal{U}_a(\alpha)} & \ghost{\mathcal{U}_a(\alpha)}& \qw & \\ } 
~~\hspace{0.5cm}\Qcircuit @C=.5em @R=.5em{
     & \qw & \qw& \multigate{1}{\mathcal{U}_y(\zeta)} & \multigate{1}{\mathcal{U}_x(\mu)} & \qw\\
    & \multigate{1}{\mathcal{U}_y(\lambda)} & \multigate{1}{\mathcal{U}_x(\sigma)} & \ghost{\mathcal{U}_a(\alpha)} & \ghost{\mathcal{U}_a(\alpha)} & \qw\\
    &  \ghost{\mathcal{U}_a(\alpha)} & \ghost{\mathcal{U}_a(\alpha)} & \qw & \qw & \qw\\ } \\
\end{align*}
\vspace{-\baselineskip}    
& 
\vspace{\baselineskip}\centering Trotter scheme $T_2$ \\[0.1em] \eqref{eq:trot2}
& \\
\midrule
\begin{align*}
\Qcircuit @C=.5em @R=.5em{
     & \multigate{1}{\mathcal{U}_{x+y}(\theta)} & \qw & \qw & \\
     & \ghost{\mathcal{U}_{x+y}(\alpha)} & \multigate{1}{\mathcal{U}_{x+y}(\theta)}  & \qw & \hspace{0.5cm} \longrightarrow \\
     & \qw & \ghost{\mathcal{U}_{x+y}(\alpha)} & \qw & \\ } 
~~\hspace{0.5cm}\Qcircuit @C=.5em @R=.5em{
     & \qw& \multigate{1}{\mathcal{U}_{x+y}(\mu)} & \qw\\
     & \multigate{1}{\mathcal{U}_{x+y}(\lambda)} & \ghost{\mathcal{U}_{x+y}(\alpha)} & \qw\\
     & \ghost{\mathcal{U}_{x+y}(\alpha)} & \qw & \qw\\ } \\
\end{align*}
\vspace{-\baselineskip}
& 
\vspace{\baselineskip}\centering Trotter scheme $T_3$ \\[0.1em] \eqref{eq:trot3}  
& \\
\midrule
\begin{align*}
\Qcircuit @C=.5em @R=.5em{
    & \multigate{1}{\mathcal{U}_x(\theta)} & \qw & \qw & \multigate{1}{\mathcal{U}_y(\theta)} & \qw & \\
     & \ghost{\mathcal{U}_a(\alpha)} & \multigate{1}{\mathcal{U}_y(\theta)} & \multigate{1}{\mathcal{U}_x(\theta)} & \ghost{\mathcal{U}_a(\alpha)} & \qw & \hspace{0.5cm} \longrightarrow \\
    & \qw & \ghost{\mathcal{U}_a(\alpha)} & \ghost{\mathcal{U}_a(\alpha)}& \qw & \qw & \\ } 
~~\hspace{0.5cm}\Qcircuit @C=.5em @R=.5em{
    & \multigate{1}{\mathcal{U}_y(\lambda)} & \qw & \qw & \multigate{1}{\mathcal{U}_x(\mu)} & \qw & \\
     & \ghost{\mathcal{U}_a(\alpha)} & \multigate{1}{\mathcal{U}_x(\sigma)} & \multigate{1}{\mathcal{U}_y(\zeta)} & \ghost{\mathcal{U}_a(\alpha)} & \qw &\\
    & \qw & \ghost{\mathcal{U}_a(\alpha)} & \ghost{\mathcal{U}_a(\alpha)}& \qw & \qw & \\ }  \\
\end{align*}
\vspace{-\baselineskip}
&
\vspace{\baselineskip}\centering Trotter scheme $T_4$ \\[0.1em]\eqref{eq:trot4}  
&  \\
\midrule
\begin{align*}
\Qcircuit @C=.5em @R=.5em{
    & \multigate{1}{\mathcal{U}_{x+y}(\theta)} & \qw & \qw & \qw & \\
     & \ghost{\mathcal{U}_{a+a}(\alpha)} & \multigate{1}{\mathcal{U}_x(\theta)} & \multigate{1}{\mathcal{U}_y(\theta)} & \qw & \hspace{0.5cm} \longrightarrow \\
    & \qw & \ghost{\mathcal{U}_a(\alpha)} & \ghost{\mathcal{U}_a(\alpha)}&  \qw & \\ } 
~~\hspace{0.5cm}\Qcircuit @C=.5em @R=.5em{
    & \qw & \qw & \multigate{1}{\mathcal{U}_{x+y}(\mu)} & \qw & \\
    & \multigate{1}{\mathcal{U}_{y}(\lambda)} & \multigate{1}{\mathcal{U}_x(\sigma)} & \ghost{\mathcal{U}_{a+a}(\alpha)} & \qw &\\
    & \ghost{\mathcal{U}_a(\alpha)} & \ghost{\mathcal{U}_a(\alpha)}& \qw & \qw & \\ }  \\
\end{align*}
\vspace{-\baselineskip}
&
\vspace{\baselineskip}\centering Trotter scheme $T_5$ \\[0.1em]\eqref{eq:trot5}   
& \\ 
\midrule
\begin{align*}
\Qcircuit @C=.5em @R=.5em{
    & \multigate{1}{\mathcal{U}_x(\theta)} & \qw  & \multigate{1}{\mathcal{U}_y(\theta)} & \qw & \\
     & \ghost{\mathcal{U}_a(\alpha)} & \multigate{1}{\mathcal{U}_{x+y}(\theta)}  & \ghost{\mathcal{U}_a(\alpha)} & \qw & \hspace{0.5cm} \longrightarrow \\
    & \qw & \ghost{\mathcal{U}_{a+a}(\alpha)} &  \qw & \qw & \\ } 
~~\hspace{0.5cm}\Qcircuit @C=.5em @R=.5em{
    & \multigate{1}{\mathcal{U}_y(\lambda)} & \qw & \multigate{1}{\mathcal{U}_x(\mu)} & \qw & \\
     & \ghost{\mathcal{U}_a(\alpha)} & \multigate{1}{\mathcal{U}_{x+y}(\mu)}  & \ghost{\mathcal{U}_a(\alpha)} & \qw &\\
    & \qw & \ghost{\mathcal{U}_{a+a}(\alpha)} & \qw & \qw & \\ }  \\
\end{align*}
\vspace{-\baselineskip}
&
\vspace{\baselineskip}\centering Trotter scheme $T_6$ \\[0.1em]\eqref{eq:trot6}  
& \\ 
\midrule
\begin{align*}
\Qcircuit @C=.5em @R=.5em{
    & \multigate{1}{\mathcal{U}_x(\theta)} & \qw & \multigate{1}{\mathcal{U}_y(\theta)}& \qw & \qw & \\
    & \ghost{\mathcal{U}_a(\alpha)} & \multigate{1}{\mathcal{U}_x(\theta)} & \ghost{\mathcal{U}_a(\alpha)} & \multigate{1}{\mathcal{U}_y(\theta)} & \qw & \hspace{0.5cm} \longrightarrow \\
    & \qw & \ghost{\mathcal{U}_a(\alpha)} & \qw & \ghost{\mathcal{U}_a(\alpha)}& \qw & \\ } 
~~\hspace{0.5cm}\Qcircuit @C=.5em @R=.5em{
     & \qw & \multigate{1}{\mathcal{U}_x(\zeta)}& \qw & \multigate{1}{\mathcal{U}_y(\mu)} & \qw\\
    & \multigate{1}{\mathcal{U}_x(\lambda)} & \ghost{\mathcal{U}_a(\alpha)} & \multigate{1}{\mathcal{U}_y(\sigma)} & \ghost{\mathcal{U}_a(\alpha)} & \qw\\
    &  \ghost{\mathcal{U}_a(\alpha)} & \qw & \ghost{\mathcal{U}_a(\alpha)}& \qw & \qw\\ } \\
\end{align*}
\vspace{-\baselineskip} 
&
\vspace{0.3\baselineskip}\centering An exact-in-principle identity used for setting up the lower bounds in the numerical tests.
& 
\\
\bottomrule
\end{tabular}
}
\caption{A list of qutrit unitary pairs $W_L(\theta, \cdots, \theta)$ and $W_R(\bm{\theta}_R)$, for which we test the approximate circuit relations through numerical minimization of the infidelity function \eqref{eq:CostFunction}. The top 6 rows correspond to the circuit reflection for different Trotter schemes \eqref{eq:trot1}-\eqref{eq:trot6}. While the unitaries in the bottom row are shown to be identical through \eqref{turnover1}, under the parameter constraint of \eqref{eq:param}, their equivalence is also tested numerically as a proxy to measure numerical deviations and limitations of optimizers.}\label{fig:trotforms}
\end{figure*}

We remark that when establishing the circuit identity \eqref{eq:ybe-qutrit} analytically, numerical validation serves as a useful and efficient strategy to verify its correctness. This step involves evaluating the following expression:
\begin{align}
\begin{split}
    \Vert &\big(\mathcal{R}_1(\alpha) \otimes \mathbf{I}_3\big)\big(\mathbf{I}_3 \otimes \mathcal{R}_2(\beta)\big)\big(\mathcal{R}_3(\gamma) \otimes \mathbf{I}_3\big) \\&-\big(\mathbf{I}_3 \otimes \mathcal{R}_4(\delta)\big)\big(\mathcal{R}_5(\epsilon) \otimes \mathbf{I}_3\big)\big(\mathbf{I}_3 \otimes \mathcal{R}_6(\zeta)\big) \Vert < \varepsilon
\end{split}
\end{align}
with a sufficiently small value of $\varepsilon$. The values for $\alpha$, $\beta$, $\gamma$, and $\delta$ are repeatedly sampled from a uniform distribution while $\epsilon$ and $\zeta$ follow from the constraint \eqref{eq:param}.

This numerical approach holds a broader range of applications and offers advantages in establishing approximate identities. While exact turnover relations may be specific to certain Hamiltonians, there are chances to develop approximate relations with tolerable levels of infidelity for a wider class of Hamiltonians. Such relations can lead to the compression of circuit depth, thus enhancing the overall fidelity of Trotter circuits running on imperfect hardware. We explore this scenario in the current section, using a concrete example of the spin-1 $XY$ Hamiltonian on three qutrits.
\begin{align}
    H_{\rm XY} = -J\sum_{i=0}^2 \left(\tilde{S}_i^x \tilde{S}_{i+1}^x + \tilde{S}_i^y \tilde{S}_{i+1}^y\right).
\label{eq:Ham}
\end{align}

The time evolution unitary of this Hamiltonian system can be written as 
\begin{align}
    e^{-i t H_{\rm XY}}
    &= e^{i Jt \sum_{i=0}^2 (\tilde{S}_i^x \otimes \tilde{S}_{i+1}^x + \tilde{S}_i^y \otimes \tilde{S}_{i+1}^y)}
    \label{eq:H_XY}
\end{align}
for which we consider a few available Trotter forms, and numerically explore if the approximate turn-over relation \eqref{eq:ybe-qutrit} holds. Specifically, for each Trotterization scheme, we evaluate $W_L(\bm{\theta}_L)$ and $W_R(\bm{\theta}_R)$, both representing unitary operators for one Trotter step and its corresponding turnover counterpart, respectively. Afterwards, we minimize the infidelity between $W_L$ and $W_R$, 
\begin{align}
\label{eq:CostFunction}
\mathcal{C}(\bm{\theta}_L, \bm{\theta}_R)= 1- \frac{1}{(3^3)^2}\left\Vert\text{tr}\,{(W_L W_R^\dagger)}\right\Vert^2,
\end{align}
by optimizing $\bm{\theta}_R$ for randomly selected values of $\bm{\theta}_L$.  
Note that the error analysis in (\ref{eq:CostFunction}) is state-independent, and the fidelity function $\frac{1}{3^6}\Vert\text{tr}\, {(W_L W_R^\dagger)}\Vert^2$ is equivalent to the mean overlap function which is a reasonable figure of merit to quantify coherent errors. 

The specific configurations of unitary pairs, $(W_L, W_R)$, that we compute to minimize the infidelity \eqref{eq:CostFunction} for, are summarized in Fig.~\ref{fig:trotforms}. Each row in the figure is related to the respective Trotter scheme for \eqref{eq:H_XY}, described below:
\begin{align}
T_1 &= \lim_{n_b\rightarrow \infty} \Big(
\mathcal{U}_{y}^{1,2}(\theta) \,
\mathcal{U}_{y}^{0,1} (\theta) \,
\mathcal{U}_{x}^{1,2}(\theta) \,
\mathcal{U}_{x}^{0,1} (\theta)
\Big)^{n_b}\label{eq:trot1}\\
T_2 &= \lim_{n_b\rightarrow \infty} \Big( 
\mathcal{U}_{y}^{1,2}(\theta) \,
\mathcal{U}_{x}^{1,2}(\theta) \,
\mathcal{U}_{y}^{0,1} (\theta) \,
\mathcal{U}_{x}^{0,1} (\theta)
\Big)^{n_b} \label{eq:trot2} \\
T_3 &= \lim_{n_b\rightarrow \infty} \Big( 
\mathcal{U}_{x+y}^{1,2}(\theta)\,
\mathcal{U}_{x+y}^{0,1}(\theta)
\Big)^{n_b}\label{eq:trot3}\\
T_4 &= \lim_{n_b\rightarrow \infty} \Big( 
\mathcal{U}_{y}^{0,1}(\theta)\,
\mathcal{U}_{x}^{1,2}(\theta)\,
\mathcal{U}_{y}^{1,2}(\theta)\,
\mathcal{U}_{x}^{0,1}(\theta)
\Big)^{n_b}\label{eq:trot4}\\ 
T_5 &= \lim_{n_b\rightarrow \infty} \Big( 
\mathcal{U}_{y}^{1,2}(\theta)\,
\mathcal{U}_{x}^{1,2}(\theta)\,
\mathcal{U}_{x+y}^{0,1}(\theta)
\Big)^{n_b}\label{eq:trot5}\\ 
T_6 &= \lim_{n_b\rightarrow \infty} \Big( 
\mathcal{U}_{y}^{0,1}(\theta)\,
\mathcal{U}_{x+y}^{1,2}(\theta)\,
\mathcal{U}_{x}^{0,1}(\theta)
\Big)^{n_b} \label{eq:trot6}
\end{align}
where $\theta = t / n_b$ and 
\begin{align}
\begin{split}
\mathcal{U}_{x}^{i,j}(\theta) &= \exp\big({-i\theta \tilde{S}^x_i\otimes\tilde{S}^x_j}\big)\\
\mathcal{U}_{y}^{i,j}(\theta) &= \exp\big({-i\theta \tilde{S}^y_i\otimes\tilde{S}^y_j}\big)\\
\mathcal{U}_{x+y}^{i,j}(\theta) &= \exp\big({-i\theta (\tilde{S}^x_i\otimes\tilde{S}^x_j + \tilde{S}^y_i\otimes\tilde{S}^y_j)}\big).
\end{split}
\end{align} 
Although the Trotter unitaries live on a one-dimensional slice of the $\bm{\theta}_L$-parameter space, i.e., $\bm{\theta}_L = (\theta, \theta, \cdots)$, it is necessary to treat all components of $\bm{\theta}_R$ independently in order to achieve a reasonably high fidelity. Therefore, we conduct numerical optimization for $\bm{\theta}_R$ with the following configuration:
\begin{align*}
    \min_{\bm{\theta}_R}\mathcal{C}(\theta, \cdots, \theta, \bm{\theta}_R) = \max_{\bm{\theta}_R} \Vert\text{tr}\,{[W_L(\theta, \cdots, \theta)\,W_R^\dagger(\bm{\theta}_R)]}\Vert^2.
\end{align*}
More generally, we consider numerical optimization using respective products of $n_b>1$ instances of $W_L$ and $W_R$.

\begin{figure*}
\begin{center}
\includegraphics[width=\textwidth]{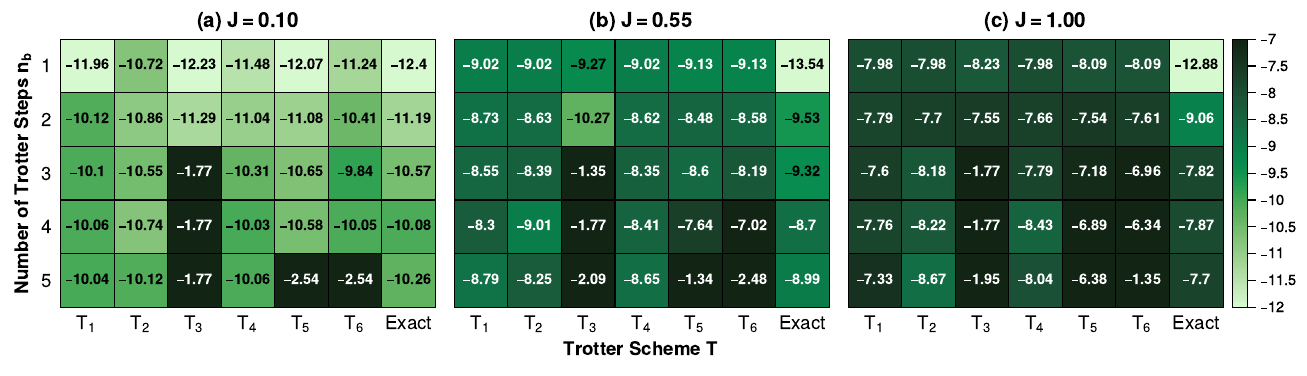}
\end{center}
\vspace{-1.5\baselineskip}
\caption{
The minimized infidelity, $\log_{10}\min_{\bm{\theta}_R}\mathcal{C}(\theta, \cdots, \theta, \bm{\theta}_R)$, is obtained through parameter optimization of $\bm{\theta}_R$ across various spin couplings $J \in \{0.1, 0.55, 1.0\}$ and Trotter schemes $T \in \{T_1, T_2, \cdots, T_6\}$. The displayed values are on the logarithmic base 10 scale. Circuit diagrams of parameterized unitaries $W_L(\theta, \cdots, \theta)$ and $W_R(\bm{\theta}_R)$ for each Trotter scheme are shown in Fig.~\ref{fig:trotforms}. The parameter optimization was performed using the BFGS algorithm. We consider the minimized infidelity to be reasonably low if it closely matches the `lower bound', which solely accounts for the numerical inaccuracy of the exact identity \eqref{eq:lbound}. 
}
\label{fig:Trotforms}
\end{figure*}

Our optimization results for infidelity are depicted in Fig.~\ref{fig:Trotforms} across different values of spin-spin coupling $J \in \{0.10, 0.55, 1.00\}$. Each panel is associated with specific $J$ values and displays the minimized infidelities on a logarithmic scale for different Trotter schemes $\{T_1, \cdots, T_6\}$, covering a range of repetition numbers $1 \leq n_b \leq 5$.
When benchmarking the infidelities between mirror-symmetric pairs of different candidate unitaries, it is necessary to establish a lower bound result that can represent practical expectations for the optimum, taking into account numerical deviations and limitations of optimizers. We achieve this through the following exact-in-principle circuit identity, 
\begin{align}
&\mathcal{U}_{y}^{1,2}(\theta)\,
\mathcal{U}_{y}^{0,1} (\theta) \,
\mathcal{U}_{x}^{1,2}(\theta) \,
\mathcal{U}_{x}^{0,1} (\theta) \notag \\ 
& ~~~~~~~~ =  
\mathcal{U}_{y}^{0,1} (\mu) \, 
\mathcal{U}_{y}^{1,2}(\sigma) \,
\mathcal{U}_{x}^{0,1} (\zeta)\,
\mathcal{U}_{x}^{1,2}(\lambda), \label{eq:lbound}
\end{align}
where $\mu$, $\sigma$, $\zeta$, and $\lambda$ should in principle be equal to $\theta$,
illustrated in the bottom row of Fig.~\ref{fig:trotforms}. It is derived from the repeated application of the exact turn-over relation \eqref{turnover1}. Therefore, its minimized infidelities should vanish ideally, i.e., $\min_{\bm{\theta}_R}\mathcal{C}(\bm{\theta}_L, \bm{\theta}_R) = 0$ for any $\bm{\theta}_L$, but are realistically sustained \textcolor{black}{through the numerical optimizer} at values ranging from $10^{-7}$ to $10^{-14}$ across different setups of $J$ and $n_b$, as depicted in Fig.~\ref{fig:Trotforms}.

In the upper six rows of Fig.~\ref{fig:trotforms}, the reflection pairs of unitaries are strategically arranged so that their repeated application within the Trotter circuit leads to a substantial reduction in the total gate count.  For instance, let us consider the Trotter scheme $T_1$ and its corresponding unitaries. The initial configuration involves 4 $\mathcal{U}_x$ and $\mathcal{U}_y$ operations for every Trotter step, summing up to a total of $4n$ gates. However, by replacing every alternating $W_L(\theta, \cdots, \theta)$ with $W_R(\mu, \sigma, \zeta, \lambda)$ at precomputed values of $\mu$, $\sigma$, $\zeta$, and $\lambda$, it becomes possible to condense $(n-1)$ gates within intermediate Trotter layers due to subsequent applications of the same unitaries, namely,
\begin{align}
    \mathcal{U}_a^{i,j}(\theta_1)\mathcal{U}_a^{i,j}(\theta_2) = \mathcal{U}_a^{i,j}(\theta_1 + \theta_2) \text{ with } a \in \{x, y\}.
\end{align} 
This results in a reduced count of $(3n+1)$ gates.

\begin{figure}
\begin{center}
\includegraphics[width=\linewidth]{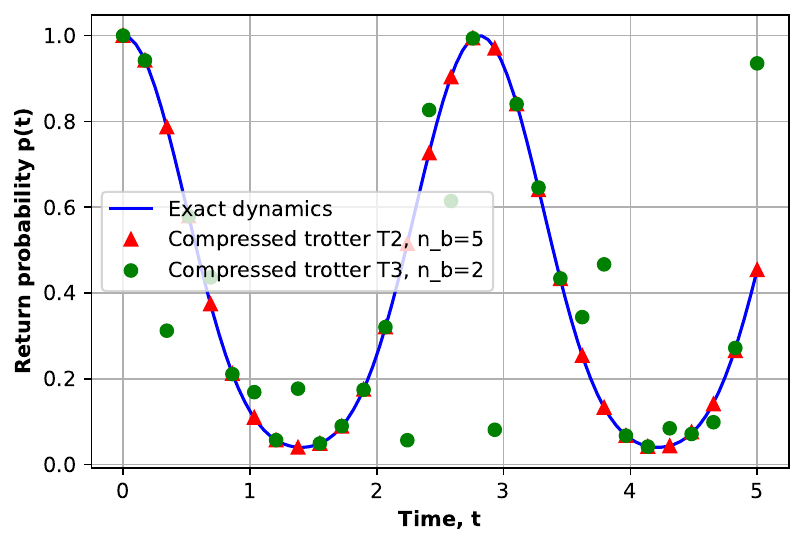}
\end{center}
\vspace{-1.5\baselineskip}
\caption{
The return probability \eqref{eq:return_prob} of the spin-1 XY model on a three-site lattice, starting and ending at the state $|202\rangle$, is displayed as a function of time within $0 \leq t < 5$. The spin coupling is set at $J=1.0$. The best Trotter scheme $(T_2, n_b = 5)$ as from Fig.~\ref{fig:Trotforms} is compared to another scheme $(T_3, n_b=2)$, with its data points represented as red and green dots.
}
\label{fig:compressed}
\end{figure}

\textcolor{black}{It is worth noting that further circuit compression is possible, but its feasibility depends on the numerical accuracy of the optimizer and the value of $J$. Consider the $T_3$ compression scheme, as illustrated in Eq.~\eqref{eq:trot3}. Continuous compression could reduce the circuit depth to $\mathcal{O}(1)$ if the turnover relation is exact. However, as observed in Fig.~\ref{fig:Trotforms}, such compression becomes possible only for $J=0.1$, $n_b=2$ and $J=0.55$, $n_b=2$, whose infidelity is even lower than that for the exact turnover identity. 
Since $n_b$ is flexible in this context, a direct comparison between two compression schemes with different $n_b$ values is inadequate. Another comparison must be conducted concerning the ``numerical performance'' of the analytically exact turn-over relations. 
Fig.~\ref{fig:compressed} provides an example of two compression schemes with different $n_b$ values tested for $J=1.0$. Despite Fig.~\ref{fig:Trotforms} indicating a minor difference in infidelities between $T_2$ with $n_b=5$ and $T_3$ with $n_b=2$, which are $10^{-8.67}$ and $10^{-7.55}$ respectively, the former is lower and the latter is higher than the numerical infidelities for the exact turnover relations at their respective $n_b$ values. This discrepancy results in a noticeable difference in actual performance under the noiseless simulation, leading to significantly better accuracy for the $T_2$ scheme at $n_b=5$.}

Hence, to attain a computational advantage with the Trotter unitary, we select $T \in \{T_1, \cdots, T_6\}$ and $1 \leq n_b \leq 5$ based on initial calculations in Fig.~\ref{fig:Trotforms}, then apply the approximate relation $W_L^{n_b} \simeq W_R^{n_b}$ for every alternate set of $n_b$ Trotter steps. While replacing $W_L^{n_b}$ with $W_R^{n_b}$ may increase the overall infidelity of the Trotter circuit, the approximation error has a negligible impact when we use $T$ and $n_b$ whose corresponding infidelity from Fig.~\ref{fig:Trotforms} stays at a level similar to the lower bound \eqref{eq:lbound}, \textcolor{black}{i.e. the last column in each panel of Fig. \ref{fig:Trotforms}}.

\textcolor{black}{Utilizing the infidelity metric outlined in \eqref{eq:CostFunction}, we derived a precise lower bound (see in Appendix. \ref{sec:lowerbound}) for the infidelity across multiple trotter steps compared to a single trotter step: $\mathcal{C}_{n_b}(\bm{\theta}_L,\bm{\theta}_R)\geq 1-(1-\mathcal{C}_1)^{n_b}$, where $\mathcal{C}_1$ is the infidelity for one Trotter step and $n_b$ denotes the exact number of trotter steps. The sharpness of this lower bound incentivizes the minimizer to actively pursue it in each optimization process involving $n_b$ trotter steps. Consequently, deviations from this lower bound signal deficiencies in the minimizer's performance. As a precautionary measure, trotter forms associated with values significantly distant from this lower bound are deliberately excluded when choosing $T\in {T_1,\ldots,T_6}$.}

\begin{figure*}
\begin{center}
\includegraphics[width=\textwidth]{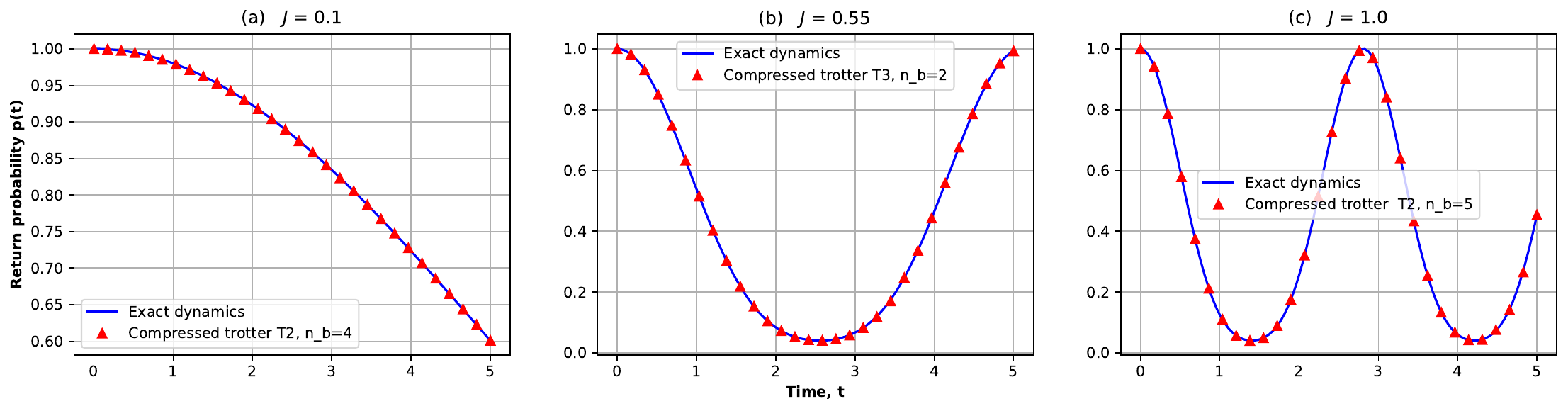}
\end{center}
\vspace{-1.5\baselineskip}
\caption{The return probability \eqref{eq:return_prob} of the spin-1 XY model on a three-site lattice, starting and ending at the state $|202\rangle$, is shown as a function of time within $0 \leq t < 5$. Each plot correspond to respective spin-couplings $J=0.1$, $0.55$, $1.0$. The dynamic simulation is represented as the blue line. We employ the Trotterization of the time-evolution operator \eqref{eq:H_XY} over 200 steps, with a corresponding step size of $\theta = 0.025$. Then we apply the circuit compression strategy detailed in Section~\ref{ssec:approx_identities} to reduce number of gates. For Trotter schemes $(T_3, n_b=2)$, $(T_2, n_b=4)$ and $(T_2, n_b=5)$, the resulting data points are indicated by small red dots. More generally, for a $(T_3, n_b=2)$ Trotter circuit with more than $n>3$ steps, the approximate count of reduced gates is $2n/3$. For a $(T_2, n_b=4)$ Trotter circuit with more than $n > 4$ steps, the approximate count of reduced gates is $\lfloor 2n/5\rfloor - 1$. For a $(T_2, n_b=5)$ Trotter circuit with more than $n > 5$ steps, the approximate count of reduced gates is $n/3 - 1$ when $n$ is a multiple of 6, and $2\lfloor n/6\rfloor$ otherwise.}
\label{fig:Trotforms2}
\end{figure*}

As a pilot application, we evaluate the returning probability of the three-qutrit system as a function of time:
\begin{align}
p(t) = \Vert\langle 202 | e^{-i t H_{\rm XY}} | 202 \rangle\Vert^2
\label{eq:return_prob}
\end{align}
with the coupling constant $J=0.1,0.55,1.0$. See the blue curve in Fig.~\ref{fig:Trotforms2} for the exact time evolution within $0 \leq t < 5$. Let us consider a Trotterization of \eqref{eq:Ham} with the step size $\theta = 0.025$, resulting in a total of $200$ steps. To employ the above circuit compression strategy, it is crucial to select an appropriate Trotter scheme based on the benchmarking outcomes in Fig.~\ref{fig:Trotforms}. For example, we find that for $J=0.55$, choosing $T=T_3$ and $n_b=2$ corresponds to a sufficiently low infidelity, $10^{-10.27}$ even less than \textcolor{black}{the numerical infidelity ($10^{-9.53}$) of the analytically exact turn-over relation}. We then impose the approximate reflection relation $W_L^{2} \simeq W_R^{2}$ sequentially, by skipping the initial $W_L$ and then substituting the following $W_L^2$ with $W_R^2$. This process repeats by skipping the subsequent $W_L$ and replacing the next $W_L^2$ with $W_R^2$, until no more $W_L^2$ remains. For $200$ Trotter blocks, the substitution is performed $66$ times, leading to the consolidation of $132$ $\mathcal{U}_{x+y}$ gates.
We show the noiseless simulation results from the approximately compressed circuits as red triangles in Fig.~\ref{fig:Trotforms2}. They demonstrate strong agreement, which is beneficial since it maintains the same level of numerical accuracy while reducing the usage of quantum resources on a qutrit-based quantum computer. All simulations for exact quantum time dynamics and its Trotterized version shown in Fig.~\ref{fig:compressed} have been performed using the QuTiP package \cite{johansson2012qutip, JOHANSSON20131234}.

\section{Conclusion and Outlook}

In this paper, we extended the discussion of searching for Yang-Baxter-like turnover relations to qudit-based quantum computing. We explored certain algebraic properties of spin-1 operators and found rigorous Yang-Baxter-like turnover relations for simple qudit models. Regarding more complex qudit models, advanced algebraic relations are challenging to resolve rigorously, but a preliminary and plausible strategy is briefly discussed, based on the conjugation relation between the spin-1 operators. Nevertheless, since large-scale quantum simulation often requires an inexact but sufficiently accurate quantum simulation, we also placed an emphasis on the numerical exploration of advanced relations for the spin-1 system. As a demonstration, we examined the spin-1 XY model and numerically explored advanced circuit turnover relations in the quantum simulation of time dynamics. In particular, we designed a pool of circuit fragment turnover pairs using various Trotterization schemes and numerically examined their fidelity to screen out potential turnover pairs that could be utilized to optimize the deep circuit corresponding to many Trotter steps in the quantum dynamics simulations. Preliminary numerical demonstrations were given on the quantum simulation of the three-qutrit XY model, where the results from our proposed numerical scheme showed great agreement with the exact curve. Remarkably, our numerical scheme can be considered as a prototype of a machine learning process to be integrated into qubit control~\cite{xu2022neural}, circuit compilation and optimization \cite{AI_for_QTech} and to improve our Yang-Baxter compiler QuYBE \cite{gulania2022quybe}. Specifically, the circuit optimization strategy can be boiled down to a combinatorial problem of searching for and performing circuit fragment turnovers in a given circuit with a layered structure, a task that can also be greatly facilitated by high-performance computing (HPC) hardware and brute-force search methods. Further studies in this direction are now underway. Our eventual goal is to develop an efficient parallel Yang-Baxter compiler such as QuYBE \cite{gulania2022quybe}, which can be used for the compression of a variety of quantum circuits, with the initial target on quantum dynamics circuits.

\section*{Acknowledgments}

This material is based upon work supported by the  U.S. Department of Energy, Office of Science, National Quantum Information Science Research Centers, Superconducting Quantum Materials and Systems Center (SQMS) under contract No. DE-AC02-07CH11359 and Next Generation Quantum Science and Engineering (Q-NEXT) under contract No. DE-AC02-06CH11357 (Basic Energy Sciences, Pacific Northwest National Laboratory (PNNL) FWP (76155)). O.O. would like to thank SQMS Algorithms Focus for helpful comments.
Y.A. acknowledges support from the U.S. Department of Energy, Office of Science, under contract DE-AC02-06CH11357 at Argonne National Laboratory. We gratefully acknowledge the computing resources provided on Bebop, a high-performance computing cluster operated by the Laboratory Computing Resource Center at Argonne National Laboratory. A.B.Ö. thanks Andrey Khesin for insightful discussions on an earlier version of the manuscript. J.K. thanks Matt Reagor for the support on this project and helpful discussions.

\appendix
\section{Derivation of the turn-over identities with $\mathcal{U}_x$, $\mathcal{U}_y$, and $\mathcal{U}_z$.}
\label{app_A}

Let $a=x$ and 
\begin{align}
    \mathcal{U}_x(\alpha) 
    &= \begin{pmatrix} 
        \mathbf{I}_3 & \mathbf{0}_3 & \mathbf{0}_3 \\
        \mathbf{0}_3 & \mathbf{I}_3- 2(\sin(\frac{\alpha}{2})\tilde{S}^x)^2 & \sin(\alpha)\tilde{S}^x \\
        \mathbf{0}_3 & -\sin(\alpha)\tilde{S}^x & \mathbf{I}_3- 2(\sin(\frac{\alpha}{2})\tilde{S}^x)^2 
    \end{pmatrix} \notag \\ \\
    &= \begin{pmatrix} 
        \mathbf{I}_3 &  \mathbf{0}_3 & \mathbf{0}_3 \\
        \mathbf{0}_3 &  \mathbf{A}_{\alpha} & \mathbf{B}_{\alpha} \\
        \mathbf{0}_3 & -\mathbf{B}_{\alpha} & \mathbf{A}_{\alpha} 
    \end{pmatrix}
\end{align}
where we denote
\begin{align}
    \mathbf{A}_{\alpha} &= \mathbf{I}_3- 2(\sin(\frac{\alpha}{2})\tilde{S}^x)^2, \\
    \mathbf{B}_{\alpha} &= \sin(\alpha)\tilde{S}^x. 
\end{align}
Now take the LHS of (\ref{eq:ybe-qutrit}) and replace $\mathcal{R}_n$ with $\mathcal{U}_x$, it is straightforward to show that
\begin{widetext}
\begin{align}
    &\big(\mathcal{U}_x(\alpha) \otimes \mathbf{I}_3\big)\big(\mathbf{I}_3 \otimes \mathcal{U}_x(\beta)\big)\big(\mathcal{U}_x(\gamma) \otimes \mathbf{I}_3\big) \notag \\
    &= \begin{pmatrix} 
        \mathbf{I}_9 &  \mathbf{0}_9 & \mathbf{0}_9 \\
        \mathbf{0}_9 &  \mathbf{A}_{\alpha}\otimes\mathbf{I}_3 & \mathbf{B}_{\alpha}\otimes\mathbf{I}_3 \\
        \mathbf{0}_9 & -\mathbf{B}_{\alpha}\otimes\mathbf{I}_3 & \mathbf{A}_{\alpha}\otimes\mathbf{I}_3
    \end{pmatrix}
    \begin{pmatrix} 
        \mathcal{U}_x(\beta) & \mathbf{0}_9 & \mathbf{0}_9 \\
        \mathbf{0}_9 & \mathcal{U}_x(\beta) & \mathbf{0}_9\\
        \mathbf{0}_9 & \mathbf{0}_9 & \mathcal{U}_x(\beta) 
    \end{pmatrix} 
    \begin{pmatrix} 
        \mathbf{I}_9 &  \mathbf{0}_9 & \mathbf{0}_9 \\
        \mathbf{0}_9 &  \mathbf{A}_{\gamma}\otimes\mathbf{I}_3 & \mathbf{B}_{\gamma}\otimes\mathbf{I}_3 \\
        \mathbf{0}_9 & -\mathbf{B}_{\gamma}\otimes\mathbf{I}_3 & \mathbf{A}_{\gamma}\otimes\mathbf{I}_3
    \end{pmatrix} \notag \\ 
    &= \begin{pmatrix} 
        \mathcal{U}_x(\beta) & \mathbf{0}_9 & \mathbf{0}_9 \\
        \mathbf{0}_9 & \mathbf{C} & \mathbf{D}\\
        \mathbf{0}_9 & -\mathbf{D} & \mathbf{C},
    \end{pmatrix}. \label{lhs_x}
\end{align}    
with
\begin{align}
    \mathbf{C} &= (\mathbf{A}_{\alpha}\otimes\mathbf{I}_3)\mathcal{U}_x(\beta)(\mathbf{A}_{\gamma}\otimes\mathbf{I}_3) - (\mathbf{B}_{\alpha}\otimes\mathbf{I}_3)\mathcal{U}_x(\beta)(\mathbf{B}_{\gamma}\otimes\mathbf{I}_3), \\ 
    \mathbf{D} &= (\mathbf{A}_{\alpha}\otimes\mathbf{I}_3)\mathcal{U}_x(\beta)(\mathbf{B}_{\gamma}\otimes\mathbf{I}_3) + (\mathbf{B}_{\alpha}\otimes\mathbf{I}_3)\mathcal{U}_x(\beta)(\mathbf{A}_{\gamma}\otimes\mathbf{I}_3)
\end{align}
\end{widetext}

Similarly, for the RHS of (\ref{eq:ybe-qutrit}), we have
\begin{widetext}
\begin{align}
    &\big(\mathbf{I}_3 \otimes \mathcal{U}_x(\delta)\big)\big(\mathcal{U}_x(\epsilon) \otimes \mathbf{I}_3\big)\big(\mathbf{I}_3 \otimes \mathcal{U}_x(\zeta)\big) \notag \\
    &= \begin{pmatrix} 
        \mathcal{U}_x(\delta) & \mathbf{0}_9 & \mathbf{0}_9 \\
        \mathbf{0}_9 & \mathcal{U}_x(\delta) & \mathbf{0}_9\\
        \mathbf{0}_9 & \mathbf{0}_9 & \mathcal{U}_x(\delta) 
    \end{pmatrix} 
    \begin{pmatrix} 
        \mathbf{I}_3 &  \mathbf{0}_3 & \mathbf{0}_3 \\
        \mathbf{0}_3 &  \mathbf{A}_{\epsilon} & \mathbf{B}_{\epsilon} \\
        \mathbf{0}_3 & -\mathbf{B}_{\epsilon} & \mathbf{A}_{\epsilon} 
    \end{pmatrix} 
    \begin{pmatrix} 
        \mathcal{U}_x(\zeta) & \mathbf{0}_9 & \mathbf{0}_9 \\
        \mathbf{0}_9 & \mathcal{U}_x(\zeta) & \mathbf{0}_9\\
        \mathbf{0}_9 & \mathbf{0}_9 & \mathcal{U}_x(\zeta) 
    \end{pmatrix} \notag \\ 
    &= \begin{pmatrix} 
        \mathcal{U}_x(\delta+\zeta) & \mathbf{0}_9 & \mathbf{0}_9 \\
        \mathbf{0}_9 & \mathcal{U}_x(\delta)\mathbf{A}_{\epsilon}\mathcal{U}_x(\zeta) & \mathcal{U}_x(\delta)\mathbf{B}_{\epsilon}\mathcal{U}_x(\zeta)\\
        \mathbf{0}_9 & -\mathcal{U}_x(\delta)\mathbf{B}_{\epsilon}\mathcal{U}_x(\zeta) & \mathcal{U}_x(\delta)\mathbf{A}_{\epsilon}\mathcal{U}_x(\zeta)
    \end{pmatrix}. \label{rhs_x}
\end{align}    
\end{widetext}
Compare (\ref{lhs_x}) with (\ref{rhs_x}), for (\ref{eq:ybe-qutrit}) to hold, the following conditions need to be satisfied,
\begin{align}
    &\mathcal{U}_x(\beta) = \mathcal{U}_x(\delta+\zeta); \label{cond_eq1} \\ 
    &\mathbf{C} = \mathcal{U}_x(\delta)\mathbf{A}_{\epsilon}\mathcal{U}_x(\zeta); \label{cond_eq2} \\
    &\mathbf{D} = \mathcal{U}_x(\delta)\mathbf{B}_{\epsilon}\mathcal{U}_x(\zeta). \label{cond_eq3} 
\end{align}
It's easy to see that (\ref{cond_eq1}) is satisfied as long as
\begin{align}
    \beta + 2k\pi = \delta+\zeta, ~~ k\in Z. \label{cond1}
\end{align}
In (\ref{cond_eq2}) 
\begin{widetext}
\begin{align}
    &(\mathbf{A}_{\alpha}\otimes\mathbf{I}_3)\mathcal{U}_x(\beta)(\mathbf{A}_{\gamma}\otimes\mathbf{I}_3) \notag \\
    &= \Big(\mathbf{I}_9- 2\sin^2(\frac{\alpha}{2}) (\tilde{S}^x)^2\otimes\mathbf{I}_3) \Big) \Big( \mathbf{I}_9 - i\sin(\beta)(\tilde{S}^x \otimes \tilde{S}^x) - 2\sin^2(\frac{\beta}{2})(\tilde{S}^x \otimes \tilde{S}^x)^2 \Big) \Big(\mathbf{I}_9- 2\sin^2(\frac{\gamma}{2}) (\tilde{S}^x)^2\otimes\mathbf{I}_3) \Big) \notag \\
    &= \mathbf{I}_9 + \bigg(\cos(\alpha)\cos(\gamma)-1\bigg)(\tilde{S}^x\otimes\mathbf{I}_3)^2 - i \cos(\alpha)\sin(\beta)\cos(\gamma) (\tilde{S}^x\otimes\tilde{S}^x) - 2 \cos(\alpha)\sin^2(\frac{\beta}{2})\cos(\gamma) (\tilde{S}^x \otimes \tilde{S}^x)^2, \\ 
    &(\mathbf{B}_{\alpha}\otimes\mathbf{I}_3)\mathcal{U}_x(\beta)(\mathbf{B}_{\gamma}\otimes\mathbf{I}_3) \notag \\
    &= \Big(\sin(\alpha)(\tilde{S}^x\otimes\mathbf{I}_3) \Big) \Big( \mathbf{I}_9 - i\sin(\beta)(\tilde{S}^x \otimes \tilde{S}^x) - 2\sin^2(\frac{\beta}{2})(\tilde{S}^x \otimes \tilde{S}^x)^2 \Big) \Big(\sin(\gamma)(\tilde{S}^x\otimes\mathbf{I}_3) \Big) \notag \\
    &= \sin(\alpha)\sin(\gamma)(\tilde{S}^x\otimes\mathbf{I}_3)^2 - i \sin(\alpha)\sin(\beta)\sin(\gamma)(\tilde{S}^x \otimes \tilde{S}^x) - 2\sin(\alpha)\sin^2(\frac{\beta}{2})\sin(\gamma)(\tilde{S}^x \otimes \tilde{S}^x)^2, \\ 
    \Rightarrow &~~ \mathbf{C} = \mathbf{I}_9 + \bigg(\cos(\alpha+\gamma)-1\bigg)(\tilde{S}^x\otimes\mathbf{I}_3)^2 - i \cos(\alpha+\gamma)\sin(\beta) (\tilde{S}^x \otimes \tilde{S}^x)
    - \cos(\alpha+\gamma) \bigg( \cos(\beta) - 1 \bigg) (\tilde{S}^x \otimes \tilde{S}^x)^2, \\ 
    & \mathcal{U}_x(\delta)\mathbf{A}_{\epsilon}\mathcal{U}_x(\zeta) \notag \\ 
    &= \Big( \mathbf{I}_9 - i\sin(\delta)(\tilde{S}^x \otimes \tilde{S}^x) - 2\sin^2(\frac{\delta}{2})(\tilde{S}^x \otimes \tilde{S}^x)^2 \Big) 
    \Big(\mathbf{I}_9- 2\sin^2(\frac{\epsilon}{2}) (\tilde{S}^x)^2\otimes\mathbf{I}_3) \Big) \notag \\
    &~~~~\times \Big( \mathbf{I}_9 - i\sin(\zeta)(\tilde{S}^x \otimes \tilde{S}^x) - 2\sin^2(\frac{\zeta}{2})(\tilde{S}^x \otimes \tilde{S}^x)^2 \Big) \notag \\ 
    &= \mathbf{I}_9 + \bigg( \cos(\epsilon) - 1 \bigg)(\tilde{S}^x\otimes\mathbf{I}_3)^2 - i \cos(\epsilon)\sin(\delta+\zeta) (\tilde{S}^x \otimes \tilde{S}^x) + \cos(\epsilon)\bigg( \cos(\delta+\zeta) - 1 \bigg)(\tilde{S}^x \otimes \tilde{S}^x)^2.
\end{align}    
Similarly, in (\ref{cond_eq3}) 
\begin{align}
    &(\mathbf{A}_{\alpha}\otimes\mathbf{I}_3)\mathcal{U}_x(\beta)(\mathbf{B}_{\gamma}\otimes\mathbf{I}_3) \notag \\
    &= \Big(\mathbf{I}_9- 2\sin^2(\frac{\alpha}{2}) (\tilde{S}^x)^2\otimes\mathbf{I}_3) \Big) \Big( \mathbf{I}_9 - i\sin(\beta)(\tilde{S}^x \otimes \tilde{S}^x) - 2\sin^2(\frac{\beta}{2})(\tilde{S}^x \otimes \tilde{S}^x)^2 \Big) \Big(\sin(\gamma)(\tilde{S}^x\otimes\mathbf{I}_3) \Big) \notag \\
    &= \cos(\alpha)\sin(\gamma)(\tilde{S}^x\otimes\mathbf{I}_3) -i \cos(\alpha)\sin(\beta)\sin(\gamma)\bigg( (\tilde{S}^x)^2 \otimes \tilde{S}^x\bigg) - 2\cos(\alpha)\sin^2(\frac{\beta}{2})\sin(\gamma)\bigg( \tilde{S}^x \otimes (\tilde{S}^x)^2 \bigg)\\
    &(\mathbf{B}_{\alpha}\otimes\mathbf{I}_3)\mathcal{U}_x(\beta)(\mathbf{A}_{\gamma}\otimes\mathbf{I}_3) \notag \\
    &= \Big(\sin(\alpha)(\tilde{S}^x\otimes\mathbf{I}_3) \Big) \Big( \mathbf{I}_9 - i\sin(\beta)(\tilde{S}^x \otimes \tilde{S}^x) - 2\sin^2(\frac{\beta}{2})(\tilde{S}^x \otimes \tilde{S}^x)^2 \Big) \Big(\mathbf{I}_9- 2\sin^2(\frac{\gamma}{2}) (\tilde{S}^x)^2\otimes\mathbf{I}_3) \Big) \notag \\
    &= \sin(\alpha)\cos(\gamma)(\tilde{S}^x\otimes\mathbf{I}_3) -i \sin(\alpha)\sin(\beta)\cos(\gamma)\bigg( (\tilde{S}^x)^2 \otimes \tilde{S}^x\bigg) - 2\sin(\alpha)\sin^2(\frac{\beta}{2})\cos(\gamma)\bigg( \tilde{S}^x \otimes (\tilde{S}^x)^2 \bigg)\\
    \Rightarrow &~~ \mathbf{D} = \sin(\alpha+\gamma)(\tilde{S}^x\otimes\mathbf{I}_3) -i \sin(\alpha+\gamma)\sin(\beta)\bigg( (\tilde{S}^x)^2 \otimes \tilde{S}^x\bigg) + \sin(\alpha+\gamma)\bigg(\cos(\beta) - 1 \bigg)\bigg( \tilde{S}^x \otimes (\tilde{S}^x)^2 \bigg)\\
    & \mathcal{U}_x(\delta)\mathbf{B}_{\epsilon}\mathcal{U}_x(\zeta) \notag \\ 
    &= \Big( \mathbf{I}_9 - i\sin(\delta)(\tilde{S}^x \otimes \tilde{S}^x) - 2\sin^2(\frac{\delta}{2})(\tilde{S}^x \otimes \tilde{S}^x)^2 \Big)     \Big(\sin(\epsilon) (\tilde{S}^x\otimes\mathbf{I}_3) \Big) \Big( \mathbf{I}_9 - i\sin(\zeta)(\tilde{S}^x \otimes \tilde{S}^x) - 2\sin^2(\frac{\zeta}{2})(\tilde{S}^x \otimes \tilde{S}^x)^2 \Big) \notag \\ 
    &= \sin(\epsilon)(\tilde{S}^x\otimes\mathbf{I}_3) -i \sin(\epsilon)\sin(\delta+\zeta)\bigg( (\tilde{S}^x)^2 \otimes \tilde{S}^x\bigg) + \sin(\epsilon)\bigg(\cos(\delta+\zeta)-1\bigg)\bigg( \tilde{S}^x \otimes (\tilde{S}^x)^2 \bigg).
\end{align}
\end{widetext}
After a term-by-term comparison, one can see that the following conditions need to be satisfied
\begin{align}
    \bigg\{\begin{array}{l}
    \alpha + \gamma = \epsilon + 2k\pi  \\
    \delta + \zeta = \beta + 2k\pi 
    \end{array},
    ~~ k\in Z. 
\end{align}
for the following equation to hold
\begin{align}
    &\big(\mathcal{U}_x(\alpha) \otimes \mathbf{I}_3\big)\big(\mathbf{I}_3 \otimes \mathcal{U}_x(\beta)\big)\big(\mathcal{U}_x(\gamma) \otimes \mathbf{I}_3\big) \notag \\ 
    &~~~~~~~~=\big(\mathbf{I}_3 \otimes \mathcal{U}_x(\delta)\big)\big(\mathcal{U}_x(\epsilon) \otimes \mathbf{I}_3\big)\big(\mathbf{I}_3 \otimes \mathcal{U}_x(\zeta)\big).
\end{align}
From (\ref{Uy}) a further proof can be obtained for $a=y$
\begin{widetext}
\begin{align}
    &\big(\mathcal{U}_y(\alpha) \otimes \mathbf{I}_3\big)\big(\mathbf{I}_3 \otimes \mathcal{U}_y(\beta)\big)\big(\mathcal{U}_y(\gamma) \otimes \mathbf{I}_3\big) \notag \\ 
    &= \bigg( (P_y\otimes P_y \otimes \mathbf{I}_3) \big(\mathcal{U}_x(\alpha) \otimes \mathbf{I}_3\big) (P_y\otimes P_y \otimes \mathbf{I}_3) \bigg) 
    \bigg( (\mathbf{I}_3 \otimes P_y\otimes P_y ) \big(\mathbf{I}_3 \otimes \mathcal{U}_x(\beta)\big) (\mathbf{I}_3 \otimes P_y\otimes P_y ) \bigg) \notag \\
    &~~~~ \times \bigg( (P_y\otimes P_y \otimes \mathbf{I}_3) \big(\mathcal{U}_x(\gamma) \otimes \mathbf{I}_3\big) (P_y\otimes P_y \otimes \mathbf{I}_3) \bigg) \notag \\
    &= (P_y\otimes P_y \otimes P_y) \big(\mathcal{U}_x(\alpha) \otimes \mathbf{I}_3\big)\big(\mathbf{I}_3 \otimes \mathcal{U}_x(\beta)\big)\big(\mathcal{U}_x(\gamma) \otimes \mathbf{I}_3\big) (P_y\otimes P_y \otimes P_y) \notag \\
    &= (P_y\otimes P_y \otimes P_y) \big(\mathbf{I}_3 \otimes \mathcal{U}_x(\delta)\big)\big(\mathcal{U}_x(\epsilon) \otimes \mathbf{I}_3\big)\big(\mathbf{I}_3 \otimes \mathcal{U}_x(\zeta)\big) (P_y\otimes P_y \otimes P_y) \notag \\
    &= \bigg( (\mathbf{I}_3 \otimes P_y\otimes P_y ) \big(\mathbf{I}_3 \otimes \mathcal{U}_x(\beta)\big) (\mathbf{I}_3 \otimes P_y\otimes P_y ) \bigg) \bigg( (P_y\otimes P_y \otimes \mathbf{I}_3) \big(\mathcal{U}_x(\alpha) \otimes \mathbf{I}_3\big) (P_y\otimes P_y \otimes \mathbf{I}_3) \bigg) \notag \\
    &~~~~ \bigg( (\mathbf{I}_3 \otimes P_y\otimes P_y ) \big(\mathbf{I}_3 \otimes \mathcal{U}_x(\beta)\big) (\mathbf{I}_3 \otimes P_y\otimes P_y ) \bigg) \notag \\
    &= \big(\mathbf{I}_3 \otimes \mathcal{U}_y(\delta)\big)\big(\mathcal{U}_y(\epsilon) \otimes \mathbf{I}_3\big)\big(\mathbf{I}_3 \otimes \mathcal{U}_y(\zeta)\big).
\end{align}    
\end{widetext}
Similar process can be followed from (\ref{Uz}) for the proof for $a=z$.

\section{Qutrit identities from qubit identities}\label{sec:qubit_qutrit_identity}

The adjoint spin-1 matrices $\tilde{S}^a$ can be seen as a $2\times 2$ block embedding of Pauli-$Y$ into a $3\times 3$ matrix, implying that the unitaries $\mathcal{U}_a$ primarily affect only two levels of the qutrit. We can build a permutation $P$ that separates the spectator levels from the levels actively involved in the unitary operations.

Taking $\tilde{S}_x$ as an example, from (\ref{sx_adj}) we have 
\begin{align*}
    P = 
    \resizebox{0.9\hsize}{!}{$\begin{psmallmatrix}
    4 & 5 & 6  & 10 & 11 & 12 & 13 & 14 & 15 & 16 & 17 & 18 & 19 & 20 & 21 & 22 & 23 & 24 \\
    5 & 6 & 4  & 11 & 12 & 15 & 19 & 20 & 16 & 21 & 22 & 10 & 13 & 14 & 17 & 23 & 24 & 18
    \end{psmallmatrix}$}
\end{align*}
which rearranges all 3-qutrit states into the following direct sum between subspaces: $|000\rangle  \oplus |00a\rangle  \oplus |0a0\rangle \oplus |0ab\rangle \oplus |a00\rangle  \oplus |a0b\rangle \oplus |ab0\rangle \oplus |abc\rangle $ with $a,b,c \in \{1,2\}$. Such permutation decomposes the 2-qutrit unitary action $\mathcal{U}_x$ into the direct sum $\mathbf{1}_{1} \oplus \exp(-i \alpha Y_1) \oplus \exp(-i \alpha Y_2)\oplus \exp(-i \alpha Y_1 \otimes Y_2)$ where $Y_{1,2}$ act on effective ``qubits" obtained from restricting qutrits onto the two-levels $|1\rangle$, $|2\rangle$. 

Applying the above permutation to the qutrit turnover relation for $\mathcal{U}_x$,
both sides of \eqref{turnover1} takes a block-diagonal form corresponding to some circuit identities on effective two-level systems.
\begin{itemize}
\item $|000\rangle$: The identity $1=1$ holds trivially.
\item $|00a\rangle$: $e^{-i \alpha Y_3} = e^{-i \delta Y_3}  e^{-i \zeta Y_3} $
\item $|0a0\rangle$: $e^{-i \alpha Y_2}e^{-i \beta Y_2}e^{-i \gamma Y_2} = e^{-i\delta Y_2} e^{-i \epsilon Y_2} e^{-i \zeta Y_2}$
\item $|a00\rangle$: $e^{-i \alpha Y_1}e^{-i \gamma Y_1} = e^{-i \epsilon Y_1}$
\item $|a0b\rangle$: $e^{-i \alpha Y_1}e^{-i \beta Y_3}e^{-i \gamma Y_1} = e^{-i \delta Y_3}e^{-i \epsilon Y_1}e^{-i \zeta Y_3}$
\item $|0ab\rangle$
\begin{align*}
    e^{-i \alpha Y_2}e^{-i \beta Y_2 \otimes Y_3}e^{-i \gamma Y_2} = e^{-i\delta Y_2 \otimes Y_3} e^{-i \epsilon Y_2} e^{-i \zeta Y_2 \otimes Y_3}
\end{align*}    
\item $|ab0\rangle$
\begin{align*}
    e^{-i\alpha Y_1 \otimes Y_2} e^{-i \beta Y_2} e^{-i \gamma Y_1 \otimes Y_2} = e^{-i \delta Y_2}e^{-i \epsilon Y_1 \otimes Y_2}e^{-i \zeta Y_2}
\end{align*}    
\item $|abc\rangle$: It returns the qubit turnover relation of \cite{Peng2022Quantum}.
\end{itemize}
The relations on $|00a\rangle$, $|0a0\rangle$, $|a00\rangle$, $|a0b\rangle$ subspaces are obviously true if the circuit parameters satisfy  \eqref{eq:param}. The $|0ab\rangle$ and $|ab0\rangle$ relations are also straightforward to verify since $[e^{-i  u Y_i}, \,e^{-i v Y_i \otimes Y_j}] = 0$. The only remaining non-trivial circuit relation comes from the subspace $|abc\rangle$, equivalent to the ``qubit" turnover relation shown in \cite{Peng2022Quantum}.

\section{Lower bound of the infidelity in multiple Trotter steps}\label{sec:lowerbound}

\textcolor{black}{In Fig. \ref{fig:Trotforms}, we reported the numerical infidelities that obtained from minimization. To evaluate the quality of the minimization which may compromise the infidelity, we can compute the lower bound of the infidelity. Similar to Eq. (\ref{eq:CostFunction}), given a Trotter scheme, for $n_b$ Trotter steps, the infidelity can be generalized to
\begin{align}
    \mathcal{C}_{n_b}(\bm{\theta}_L, \bm{\theta}_R) &= 1- \frac{1}{(3^3)^{2n_b}}\left\Vert\text{tr}\,\left[ (W_{L})^{n_b} (W_{R}^\dagger)^{n_b}\right] \right\Vert^2. \label{eq:c_nb}
\end{align}
If $W_{L}W_{R}^\dagger$ is positive semidefinite (which can be generally true when $W_{L}W_{R}^\dagger \rightarrow I$), we then have 
\begin{align}
    \text{tr}\,\left[ (W_{L})^{n_b} (W_{R}^\dagger)^{n_b}\right] &= \text{tr}\,\left[ (W_{L}W_{R}^\dagger)^{n_b} \right] \notag \\
    &\le \left[ \text{tr}\,(W_{L}W_{R}^\dagger) \right]^{n_b},
\end{align}
and the lower bound of (\ref{eq:c_nb}) can be expressed as
\begin{align}
    \mathcal{C}_{n_b} 
    &\ge 1 - \left[ 
    \frac{1}{(3^3)^2} \left\Vert \text{tr}\,(W_{L}W_{R}^\dagger) \right\Vert^2
    \right]^{n_b} \notag \\
    &= 1 - ( 1 - \mathcal{C}_1 )^{n_b}.
\end{align}
Here $\mathcal{C}_1$ is the infidelity for one Trotter step. Now based on the computed lower bounds, we can evaluate the minimized infidelities reported in Fig. \ref{fig:Trotforms}. In particular, if the minimized infidelity is far away from the computed lower bound such as T3 ($n_b\ge3$), the minimization become deficient.}

\bibliographystyle{plain}
\bibliography{refs}

\end{document}